\documentclass[sn-mathphys,Numbered,icol]{sn-jnl}

\usepackage{graphicx}%
\usepackage{multirow}%
\usepackage{amsmath,amssymb,amsfonts}%
\usepackage{amsthm}%
\usepackage{mathrsfs}%
\usepackage[title]{appendix}%
\usepackage{xcolor}%
\usepackage{textcomp}%
\usepackage{manyfoot}%
\usepackage{booktabs}%
\usepackage{algorithm}%
\usepackage{algorithmicx}%
\usepackage{algpseudocode}%
\usepackage{listings}%

\usepackage{tikz,xcolor,hyperref}

\definecolor{lime}{HTML}{A6CE39}
\DeclareRobustCommand{\orcidicon}{
	\begin{tikzpicture}
	\draw[lime, fill=lime] (0,0) 
	circle [radius=0.16] 
	node[white] {{\fontfamily{qag}\selectfont \tiny ID}};
	\draw[white, fill=white] (-0.0625,0.095) 
	circle [radius=0.007];
	\end{tikzpicture}
	\hspace{-2mm}
}

\foreach \x in {A, ..., Z}{\expandafter\xdef\csname orcid\x\endcsname{\noexpand\href{https://orcid.org/\csname orcidauthor\x\endcsname}
			{\noexpand\orcidicon}}
}

\begin{document}

\title{Antifragile control systems in neuronal processing: A sensorimotor perspective}


\author[ ]{\fnm{Cristian} \sur{Axenie}\orcidA{}}
\email{cristian.axenie@th-nuernberg.de}

\affil[ ]{\orgdiv{Department of Computer Science and Center for Artificial Intelligence} 
\orgname{Technische Hochschule Nürnberg Georg Simon Ohm, } \orgaddress{\street{Keßlerplatz 12} \postcode{90489} \city{Nürnberg}, \country{Germany}}}


\abstract{
The stability--robustness--resilience--adaptiveness continuum in neuronal processing follows a hierarchical structure that explains interactions and information processing among the different time scales. Interestingly, using "canonical" neuronal computational circuits, such as Homeostatic Activity Regulation, Winner-Take-All, and Hebbian Temporal Correlation Learning, one can extend the behaviour spectrum towards antifragility. Cast already in both probability theory and dynamical systems, antifragility can explain and define the interesting interplay among neural circuits, found, for instance, in sensorimotor control in the face of uncertainty and volatility. This perspective proposes a new framework to analyse and describe closed-loop neuronal processing using principles of antifragility, targeting sensorimotor control. Our objective is two-fold. First, we introduce antifragile control as a conceptual framework to quantify closed-loop neuronal network behaviours that gain from uncertainty and volatility. Second, we introduce neuronal network design principles, opening the path to neuromorphic implementations and transfer to technical systems.}


\maketitle

\section{Preamble}

Feedback control theory offers the most versatile and comprehensive set of analysis and synthesis methodologies for dynamical systems under prescribed dynamics. Interestingly, in technical systems, this framework is instantiated as a set of qualitative mechanistic explanations and means of attaining prescribed dynamics in the presence of \textit{uncertainty} and \textit{volatility}. In contrast, in biological systems, \textit{feedback control theory} is used as a qualitative metaphor to generate hypotheses on maintaining function under the impact of \textit{un-modelled dynamics} and \textit{perturbations}. 

\textit{Neuronal processing} offers a very complex and subtle canonical system that exposes a large repertoire of dynamics. Under the impact of perturbations that can take place over \textit{timescales} from milliseconds to months or even years, \textit{neural networks} need to keep their \textit{stability}. It seems paradoxical, but the only way neural networks can stay stable is if they are excitable, able to \textit{adapt} their response (and structure) in reaction to outside stimuli, and able to respond appropriately. Neural networks, like other \textit{biological systems}, exhibit a broad \textit{spectrum of behaviours} concerning modifications at the neuron-level excitability that regulate and improve their functionality, absorbing a broad variety of molecular and cellular parameter changes while preserving their spiking functionality. However, the spectrum of behaviours is typically built with a strong reference to the basic state of stability.

In this perspective, we refer to the stability of a dynamical system as the tendency to return to the initial, steady-state, balanced relations between the components that operate within it (i.e., post-spike refractory period and cell re-polarization) upon the elimination of an input disturbance. Advancing on the spectrum, we hereby define robustness as the ability of the system to endure disturbances or variations in its input. Tightly coupled, we define resilience as the ability of a system to regain its equilibrium once it experiences a variety of variations in parameters. In this case, despite these modifications, the system reacts by modifying its internal states to preserve its general function. For instance, consider the excitability of individual neurons or neural networks that can withstand alterations in ionic channel expression, frequency of stimulation, temperature, salinity, and pH. Neurons are plastic, capable of adapting their behaviour when faced with a new task, i.e., input pattern change concerning their frequency-current characteristic. This describes the next behaviour in the spectrum, namely \textit{adaptation}. This behaviour is defined as the set of immediate adjustments in a cell or system (i.e., a neural network) triggered by a stimulus that persists.

Hitherto, embedded in the formalism of feedback control theory, we extend the behavioural spectrum of neural processing. We introduce \textit{antifragility} as a new member of the spectrum that goes beyond \textit{robustness}, \textit{resilience}, and \textit{adaptation}. It leverages the multiple time scales neural processing unfolds upon to capture how single neurons and neural networks not only absorb and react to changes but gain from volatile disturbances and uncertainty. We provide a new perspective on how single-cell, within neural networks, and between neural network dynamics work in concert, in closed-loop, to build the capacity to anticipate changes in their input and, hence, gain from the inherent uncertainty within.

The role of this \textit{perspective\footnote{Research designed and developed by the interdisciplinary \href{http://antifragility.science/}{Applied Antifragility Research Group}}} is to encourage the community to discuss the extension of the neural processing behaviour spectrum and consider antifragility a "first-class citizen" beyond what the dynamical systems framework has already postulated. Feedback control theory comes as a very useful tool to support this initiative, offering both intuitive and formal support to introduce antifragility. We are taking our first steps towards the unified neuronal behaviour spectrum, so we will only target, from this perspective, neuronal sensorimotor control. We build the perspective progressively, from evidence of resilience and robustness to uncertainty coding in neuronal processing up to sensorimotor models and their inherent uncertainty. We conclude by embedding sensorimotor control in the antifragile feedback control framework, and, finally, suggesting a variety of research directions stemming from this new perspective.

\section{Neural correlates and representations of uncertainty}

From the single neuron upstream, \textit{uncertainty} and \textit{volatility} are explicitly encoded in various ways. Framing uncertainty as an inherent property of the environment, central in internal models of decision-making and learning the study of \cite{payzan2013neural} used fMRI experiments to model the encoding of uncertainty in various locations in the brain. The work defined using Bayesian tools pathways that modulate noradrenergic representations of uncertainty (i.e. rick, estimation uncertainty, unexpected uncertainty) in value-based decision-making. Extending this direction, volatility was defined as a nonlinear combination of onset, duration and amplitude of external signal disturbance. This view stems from the excellent study on imprecise neural computation as a source of adaptive behaviour in volatile environments of \cite{findling2021imprecise}. The study defined the statistical behaviour in the presence of varying and fixed volatility models leveraging: 1) high-order inferences about the environmental volatility, 2) neural computations that derive posterior beliefs, and 3) computational imprecisions that scale with the magnitude of changes in internal representations.

In a broader context it has been shown that novelty and uncertainty work in concert to regulate exploration vs exploitation \cite{cockburn2021novelty} in neural processing of risk and ambiguity \cite{wu2021better} by either employing a prediction error of uncertainty driving sensory learning \cite{iglesias2021cholinergic} or simply the adaptive learning under expected and unexpected uncertainty \cite{soltani2019adaptive}. Interestingly, a similar scheme was found across multiple studies where neural processing and learning were modulated through uncertainty \cite{grossman2022serotonin} using explicit neural representations of uncertainty and ambiguity \cite{bach2011known} for multiple tasks, of which serial decision uncertainty under probabilistic representation \cite{van2019probabilistic} was representative for sensorimotor control. Overall, it seems that neural coding of uncertainty and probability \cite{ma2014neural} plays the central role, not only in obtaining stable and robust representations of sensory and motor streams but also in driving the neural organization of uncertainty estimates \cite{bach2012knowing}.

These observations pave the way to our control theoretic framework where we embed neuronal processing. Combined experimental and modelling studies, such as of \cite{muller2019control} already characterized mechanisms for the control of entropy in neural models of environmental representations with a multivariate approach where multiple temporal and spatial scales were calibrated. This perspective emerged from a previous study which only handled partially the closed-loop perspective, where the role of dopamine was discussed in modulated affordances for active inference \cite{friston2012dopamine} or the modulation of attention under uncertainty modelled as a free-energy problem \cite{feldman2010attention}. The rather vast landscape of perspectives and models in the literature makes it hard to embed all research paths in our framework but we revamp those concepts and elements which motivate our endeavour and the need to assess the varieties of uncertainty in decision-making \cite{bland2012different}, the impact uncertainty has upon cognitive control \cite{mushtaq2011uncertainty}, and, of course, the uncertainty types and representations mediating attention, learning, and decision-making \cite{monosov2020outcome}.

Finally, from a closed-loop system perspective, the rather new perspectives on reward-based reinforcement learning and gaining under uncertainty and volatility pave the way for extending the stable-robust-resilient-adaptive behaviour spectrum. Well-known is the fact that reinforcement learning is gaining while finding optimal balance exploitation vs exploration, but this takes another dimension when uncertainty is represented under neuromodulation and attention \cite{angela2005uncertainty}. Here, one typically considers explicit neural signals reflecting uncertainty \cite{schultz2008explicit} in the closed reinforcement learning loop and the fact that the sensorimotor system itself contains components supporting perception under uncertainty, for instance through the thalamocortical excitability modulation  \cite{kosciessa2021thalamocortical}.

\section{Robustness and resilience in neuronal processing}

Neuronal networks must be stable to persist the learned relationships between the various sensory and motor streams they are modulated by and their internal states. Paradoxically, such systems can only remain stable, from a dynamical systems perspective, if they are excitable, able to adapt their behaviour in reaction to outside stimuli, and able to withstand those changes \cite{cannon1929organization}. At the same time, neural networks are flexible, so in a way, they are stable \cite{musslick2019stability}; actually, a network's ability to accommodate slight variations in its parameters, operating variables, and state variables determines how stable it is overall \cite{holling1973resilience}. Finally, a neuronal network is plastic and able to adapt its behaviour in response to new input configurations, tasks, and noise patterns within its components and driving variables \cite{braun2015unforeseen}.

Our goal is to shed new light and extend the stability-robustness-resilience-adaptation spectrum of neuronal processing by virtue of a novel feedback control-theoretic framework, namely antifragile feedback control systems. We build upon both theoretical dynamical systems analysis \cite{krakovska2021resilience} as well as biophysical evidence and modelling of neuronal networks \cite{marom2023biophysical} and propose a spectrum of behaviours where robustness, resilience, and adaptation are members of a broader \textit{stability-antifragility continuum}. As coined in the work of Nasim Taleb \cite{taleb2012Antifragile,taleb2013mathematical}, a dynamical system's propensity to benefit from \textit{unpredictability, volatility, and uncertainty} in contrast to what fragility would incur is defined as \textit{antifragility}. The response of an antifragile dynamical system to perturbations is \textit{beyond robust and resilient}, to the point where \textit{stressors} can improve the system's stability and response by contributing a significant \textit{anticipation component}.

Neuronal robustness and resilience unfold across scales \cite{marom2023biophysical}, as seen through the lens of feedback control systems. In a very nice mechanistic build-up, the authors introduce a behaviour spectrum, from stability to adaptation through robustness and resilience. This is the starting point of our creative exercise to extend the behaviour range with a new member, namely antifragility. This initiative is not only theoretical; rather, it builds upon previous research on the underlying mechanisms of robustness and resilience in neural systems. The current stability-resilience spectrum proposes that each behaviour of single neurons and neural networks can seamlessly describe the transition between multiple mechanisms to sustain function in the face of disruptions occurring on timelines spanning multiple orders of magnitude due to the interleaving of distinct timescales, as suggested by the excellent framework of the comprehensive theory of adaptive variation \cite{meyers2002fighting}.

To extend the spectrum, we build on the key insights from the dynamical systems analysis, as formalized in both \cite{krakovska2021resilience,meyer2015dynamical}. We therein consider that: 1) resilience to recurrent state variable alterations correlates with resilience to changes in parameters due to the critical slowing down phenomenon (i.e., homeostatic activity regulation (HAR) in single neurons); 2) neuronal networks provide immunity to state variable fluctuations that may be different from those that promote recovery from them (i.e., competition and cooperation in the population of neurons encoding single sensory and motor streams through Winner-Take-All (WTA)); and 3) recovery rates matter for biological resilience (i.e., temporal correlation learning through Hebbian Learning (HL) variations).

But the question is: how can we embed strategies and recovery mechanisms in a mathematical framework for resilience dynamics under uncertainty while still ensuring stable robustness? In a very interesting study, \cite{lara2018mathematical} provided a definition of resilience as a form of controllability for whole random processes (regimes), whereas the state values must belong to an acceptable subset of the state set. To achieve this behaviour, a mix of positive and negative feedback loops is needed to sustain this internal control signal propagation under both functional and structural changes \cite{hebbar2022interplay}. In our endeavour, the dynamical analysis of neuronal processing systems needs to be complemented by formal measures of each behaviour in the stability-resilience spectrum. In this respect, the study of \cite{bramson2010formal} demonstrated that systems exhibiting feedback, nonlinearity, heterogeneity, and path dependencies need to be captured in a unified mathematical object. The study considered the Markov model framework provided above to establish formal definitions of several concepts in the dynamical systems behaviour spectrum: robust, reliable, sustainable, resilient, recoverable, stable, and static, as well as their counterparts: susceptible, vulnerable, and fragile. This versatile framework demonstrates again the need to extend the behaviour spectrum in the realm of dynamical systems with prescribed dynamics. Yet, the treatment in the study was oriented toward graph theory and statistics, whereas the practical closed-loop dynamics were not central. This motivates our perspective to stand out as an endeavour towards feedback control systems to describe antifragility.

From a quantitative angle, there is research on wrapping probabilistic techniques for assessing the resilience of complex dynamical systems in feedback control loops. As emphasized in the study of  \cite{balchanos2012probabilistic}, a feedback controller is responsible for system performance recovery through the application of different reconfiguration strategies and strategic activation of necessary redundancy. Hence, uncertainty and volatility effects on a system's operation are captured by disturbance factors. These observations are immediately valid in the biological systems realm. In this direction, the study of \cite{arnoldi2016resilience} on resilience, reactivity, and variability redefines the spectrum of stable, robust, resilient, and adaptive behaviours within the framework of geometric eigenvector-based metrics. Parameterized using the eigenvalues distance from equilibrium points, these metrics capture time-scale separation and order reduction through eigenvector motion parameters. 

Using the same framing of robustness in the geometrical framework of response shape in flow networks, the work of \cite{ay2007geometric} introduces structural robustness as the core of capturing the causal contribution of each system component to the network's robustness. This perspective is amenable to neuronal processing as functional redundancy plays a fundamental role in the robustness and resilience of the system's response and, as we will see later, in its antifragility.
Finally, the comprehensive work on resilience in dynamical systems by \cite{krakovska2021resilience}, compiles a formal set of metrics and analysis mechanisms to describe the ability of a dynamical system to absorb changes in state variables, driving variables, and parameters and still persist. The framework was formalized along a very concise set of metrics such as return time (reaching time, proportional to the reciprocal of the eigenvalue with the largest real part of the system linearization), reactivity (the maximum instantaneous rate at which an asymptotically stable linear system responds if initial conditions are away from the origin), and intrinsic stochasticity.

This section has set the stage for the "tour de force" we perform in extending the behavioural spectrum of neuronal processing towards antifragility. We now have all the core concepts and framing defined and ready to introduce the specifics of our canonical system, namely the sensorimotor system.

\section{Robust sensorimotor control under uncertainty}

The most appropriate way to conceptualize sensorimotor control is as a highly elaborate and intricate process that involves thousands of ensembles of peripheral sensory data processed by a network of neurons, interneurons, and central nervous system regions. This interconnected structure then uses an equally sophisticated network of pathways and neural networks to stimulate the muscles and generate coordinated movements.

Typically formulated in the Bayesian framework for decision-making, its core is about constructing a representation of the state of the world used subsequently to make decisions based on describing uncertainty as probability distributions. Sensorimotor processes are plagued by uncertainty, which stems from a variety of causes including sensory and motor noise, as well as environmental uncertainty \citep{orban2011representations}. However, it remains unclear if uncertainty is task-dependent, only at the decision-making level, or completely Bayesian, across the whole perceptual machinery \cite{koblinger2021representations,kording2004bayesian}.

Biases and optimality in cognitive decision-making form a strong reference frame when considering robustness and resilience, especially when experimental data demonstrate that motor strategy selection comes close to maximizing expected gain \cite{trommershauser2009biases}. However, this is only one pathway, whereas decision-making and movement planning are better represented in statistical decision theory. Here, economic decision-making tasks generally do not optimize expected benefits and frequently underestimate the probability of rare events \cite{trommershauser2008decision}.

It seems that the sensorimotor system inherently robustifies its output on the long tails \cite{taleb2020statistical} by integrating sensory and motor information, each with different noise properties (i.e., reliability), in a way that minimizes the uncertainty in the overall estimate \cite{van2002role}. Such observations are even more well captured in the framework of affordances discovery through perception and action \cite{chavez2016discovering}. More precisely, sensorimotor task performance is maximized by adapting the dynamics of the system under physical and computational constraints \cite{ogawa2006adaptive}.

Overall, the capacity to absorb changes in its parameters and input and recover from volatile disruptions remains in the realm of statistical optimal perception \cite{fiser2010statistically}. Even when formulated as feedback control loops, the computational models of sensorimotor integration still exploit the closed-loop dynamics to cope with noise in the input, disturbances and (task) structure changes \cite{ghahramani1997computational}.

It seems that the stability-robustness-resilience-adaptiveness continuum in sensorimotor control also follows a hierarchical structure \cite{nagata1994hierarchical} that explains the interactions among the different time scales of sensory integration, motor plan generation, and disturbance compensation, overall under a clear impact of coordinate transformation uncertainty \cite{schlicht2007impact}. Multiple explanations and models have been used to capture aspects of the role of uncertainty in neural coding and computation underlying sensorimotor control \cite{knill2004bayesian} where the Bayesian approaches to sensory integration \cite{berniker2011bayesian} seem to dominate. Yet, none of the approaches captures the consequences of fat tails \cite{taleb2020statistical}, statistical moments and volatility, and the geometry of the system's response over time \cite{taleb2013mathematical} while learning under uncertainty \cite{topel2023expecting}.

\section{Antifragility in sensorimotor processing}

Antifragility was introduced as a very versatile and powerful framework to describe a system's behaviour in the face of randomness, uncertainty, and volatility by Taleb \cite{taleb2012Antifragile,taleb2013mathematical}. In its initial form, the mathematical treatment was centred on the probabilistic aspects of a system's behaviours under uncertainty and volatility while describing a new fragile--antifragile spectrum of responses. Our endeavour starts here and transfers antifragility principles into the realm of the dynamical system by adding the time component. This is a crucial ingredient that allows us to define, measure, and integrate antifragility principles in control theory. This interdisciplinary research is currently carried out by the \href{http://antifragility.science/}{Applied Antifragility Group}.

For an introduction to applied antifragility, we invite the reader to consult the work in \cite{axenie2023antifragility}, where we classify antifragility across scales: from intrinsic antifragility (i.e., describing the dynamical system's intrinsic temporal dynamics), to inherited antifragility (i.e., determined by the system's local interactions with other systems), and up to induced antifragility (i.e., prescribed dynamics in a closed-loop feedback control paradigm).
The main ingredients of the antifragile control theory were refined across multiple instantiations \cite{10345540, axenie2023antifragile,axenie2022antifragile} and comprise:
\begin{itemize}
    \item time-scale separation 
    \item redundant overcompensation
    \item variable structure and attractor dynamics
\end{itemize}

Interestingly, antifragility principles found applicability even beyond our control theoretic formulation towards: antifragility based on constraint-satisfaction in networks \cite{pineda2019novel}, robustness and fragility based on feedback dynamics \cite{kwon2008quantitative}, antifragility in systems of systems \cite{johnson2013antifragility}, and antifragility criteria for modelling dynamical systems \cite{de2020antifragility}. This broad landscape consolidates our perspective and frames it better in the dynamical systems domain.

To facilitate the conceptual work we develop from this perspective, we anchor the explanations in a limited but representative set of neural sensorimotor control models. Initial studies framing sensorimotor neuronal processing in control theory covered classical aspects of dynamical systems controllability and stabilization \cite{levin1993control}, observability and identification \cite{levin1996control}, and closed-loop feedback control \cite{narendra1996neural}. In our current perspective, we analyse the characteristics of representative sensorimotor models through the lens of antifragility and its three main ingredients. Additionally, to adapt the framework to neuronal processing, we limit ourselves to the analysis of only three main dynamical (computational) elements: HAR, WTA, and HL, respectively. As example systems, we consider the unsupervised learning of sensorimotor relations network of \cite{cook2010unsupervised}, the self-organising sigma-pi sensorimotor network of \cite{weber2007self}, the interacting basis function network of \cite{firouzi2014flexible}, and the sensorimotor correlation learning network of \cite{axenie2016self}. Our choice of models and the computational mechanisms is motivated by multiple recent computational and experimental studies emphasizing the major role of the "canonical" computational mechanisms (i.e., HAR, WTA, HL) in neuronal processing behaviours.

\section{Antifragility levels in neural processing}

Our perspective unfolds, in the current section, a creative exercise that embeds sensorimotor control mechanisms in the antifragility framework. Considering a selection of relevant models and the framework of the perspective, we provide insights on antifragile theory analysis and design. As reference models, we consider the sensorimotor relations network of \cite{cook2010unsupervised}, the self-organising sigma-pi sensorimotor network of \cite{weber2007self}, the interacting basis function network of \cite{firouzi2014flexible}, and the sensorimotor correlation learning network of \cite{axenie2016self}. The levelled approach to introduce antifragility takes into account that the "canonical" neuronal circuits we use operate on quite different time scales. The WTA dynamics operate on a short time scale, allowing the neuronal network to converge quickly. HAR and HL operate on a much longer time scale, averaging over a much larger sample of inputs. 

\subsection{Intrinsic antifragility}

Bottom-up, we argue that the implementation of the antifragile controller is performed through the HAR dynamics. This acts as a Proportional Derivative (PD) controller tuned in response to changes in behavioural states, experience, and learning \cite{ruggiero2021mitochondria}. The study shows that mitochondria are key mediators of HAR. The release of synaptic vesicles and intracellular calcium concentration were provided as examples of neuronal variables that are known to be regulated by mitochondria. Using fundamental ideas from control theory, the study developed a classification scheme for potential homeostatic machinery parts that stabilize firing rates. However, the physiological variables underlying this process and their cellular underpinnings or neural network components are still not well identified.

\begin{figure}
    \centering
    \includegraphics[width=0.8\textwidth]{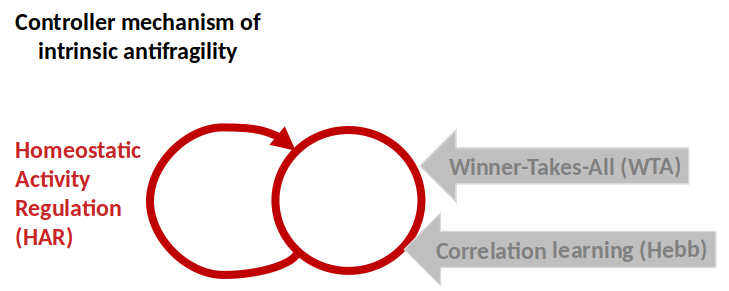}
    \caption{Intrinsic antifragility core mechanism based on Homeostatic Activity Regulation (HAR). The other within-population dynamics (i.e. Winner-Takes-All) and between-population (i.e. Correlation Learning) impact local dynamics, such that the neuron activation is a superposition of multiple sources with inherent own noise, distribution, and reliability properties.}
    \label{fig:intrinsic-antifragility-core}
\end{figure}

From the computational side, the model of \cite{cook2010unsupervised} demonstrates that HAR ensures the adaptation to local processing at the neuron level while preserving the consistent balance of \textit{time scale separation} among WTA, HL, and HAR (see Figure~\ref{fig:intrinsic-antifragility-core}). This is confirmed also in the model of \cite{axenie2016self}, where the HAR circuitry ensures that each neuron is active roughly a given proportion of the time, making sure that every neuron is active, and that each neuron is used in moderation.
The antifragile feedback control loop for intrinsic antifragility builds around the basic control mechanism in Figure~\ref{fig:intrinsic-antifragility-core} and describes neuron-level processing. The closed-loop feedback mechanism is here centred on the HAR dynamics implementing, together with the neuron model, the controller that compensates for disturbances and uncertainty when trying to produce the response pattern close to the prescribed target response pattern.
\begin{figure}
    \centering
    \includegraphics[width=0.8\linewidth]{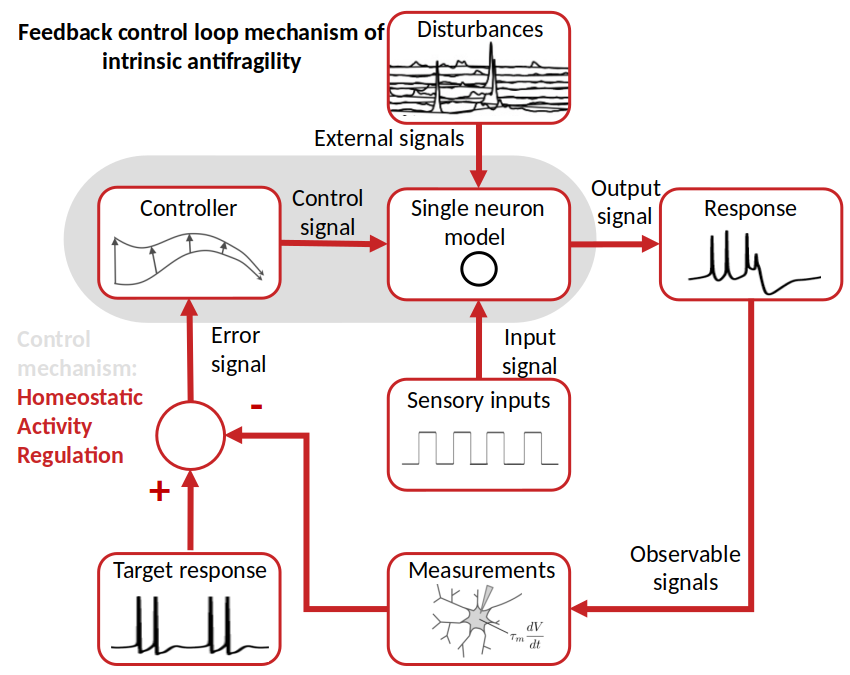}
    \caption{Intrinsic antifragility: feedback control loop for intrinsic antifragility components. The Controller implements computational mechanisms specific to Homeostatic Activity Regulation (HAR) and its interaction with the single-neuron model dynamics.}
    \label{fig:intrinsic-antifragility}
\end{figure}
Intrinsic antifragility captures those dynamics of the single neurons which are determined by the physical, chemical, and electrical properties of the neuron type and function.

\subsection{Inherited antifragility}

\begin{figure}
    \centering
    \includegraphics[width=0.8\linewidth]{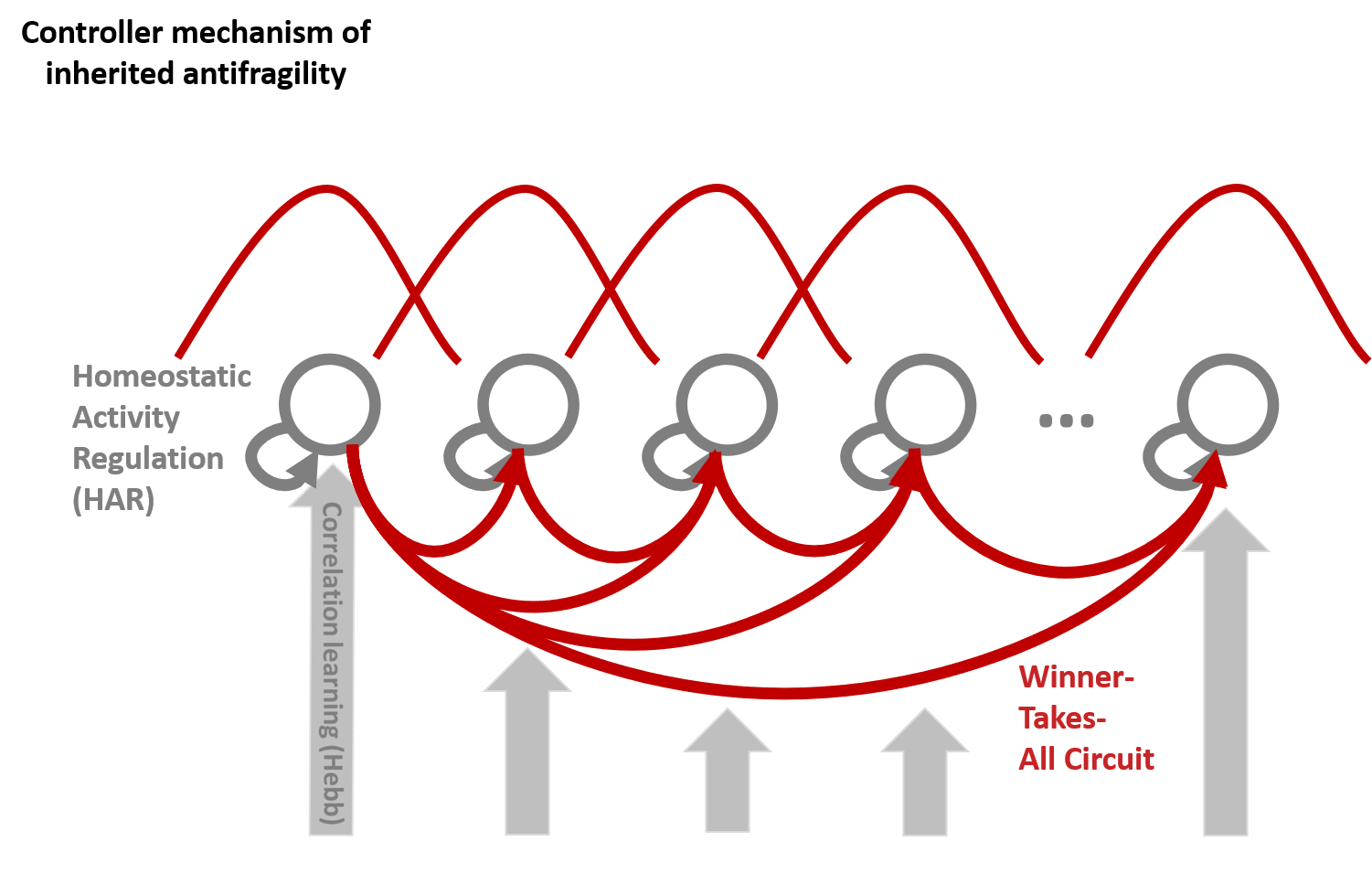}
    \caption{Inherited antifragility core based on Winner-Take-All within-population dynamics. Tuning curve modulation determines capacity building and sensorimotor input data distribution used in the competition and cooperation mechanisms acting between the neurons, whose activity is HAR controlled.}
    \label{fig:inherited-antifragility-core}
\end{figure}

\begin{figure}[H]
    \centering
    \includegraphics[width=0.8\linewidth]{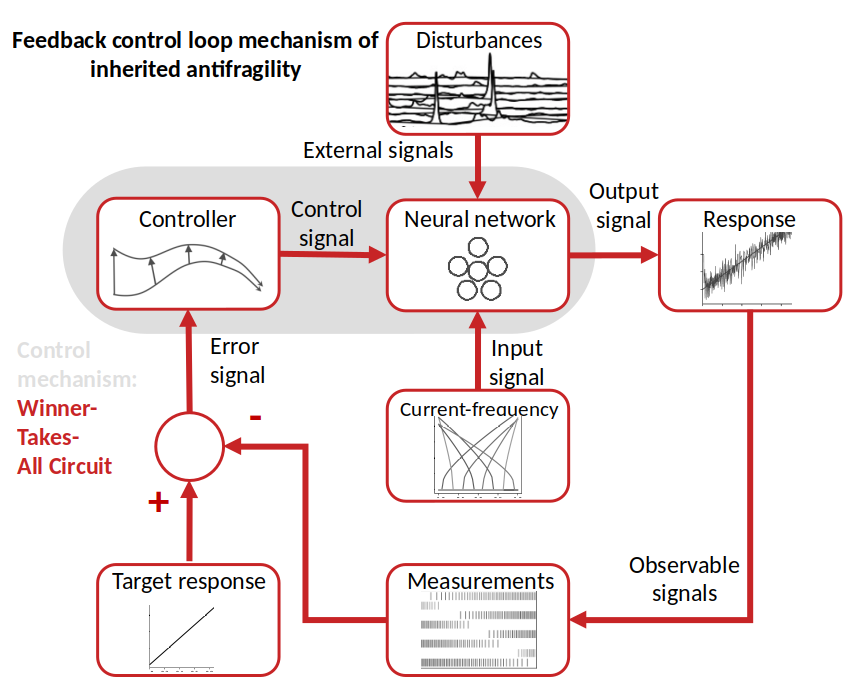}
    \caption{Inherited antifragility: feedback control loop for inherited antifragility components. The Controller implements computational mechanisms specific to neural population competition and cooperation (i.e. Winner-Take-All circuit) and its interaction within a neural population.}
    \label{fig:inherited-antifragility}
\end{figure}

Considering the neuronal network level, we hereby introduce inherited antifragility. At this point, neuronal population dynamics (e.g. competition and cooperation, depicted in Figure~\ref{fig:inherited-antifragility-core}) dictate the shape of the closed-loop dynamics. The superimposed effect of HAR and WTA now dictates the within-population self-organization, through time-scale harmonization and redundant overcompensation through a judicious modulation of neurons' tuning curves.
In this context, the study of \cite{kim2022dynamical} sheds light on the dynamical origin of WTA competition within neuronal populations. Using a network of the hippocampus dentate gyrus, the study examines the dynamical origin of WTA which results in sparse activation of the granule cell clusters. Accordingly, WTA dynamics arise from a competition between the inhibitory cells' feedback and the firing activity inside each neuronal cluster. The biophysical results are further confirmed by the computational sensorimotor control studies of \cite{cook2010unsupervised}, \cite{weber2007self}, \cite{firouzi2014flexible}, and \cite{axenie2016self}. 
Herein, WTA circuitry is responsible for balancing time scale separation, by ensuring fast convergence but also by building capacity (i.e. redundant overcompensation) of the neuronal representation of the sensorimotor streams. Complementing slower HAR dynamics, WTA implements at the neuronal population level the antifragile controller responsible for reproducing the prescribed response under the effect of disturbances, local HAR dynamics, and the impact of the inhibitory and excitatory within-population dynamics (see Figure~\ref{fig:inherited-antifragility}).
Inherited antifragility describes the population-level compound dynamics of time-scale harmonization and modulating action of competition and cooperation to represent the sensorimotor input signals.

\subsection{Induced antifragility}

At the highest level, the interplay of both fast and slow timescales of the WTA and HAR circuits is combined with temporal correlation learning exhibited between neural populations \cite{cook2010unsupervised,firouzi2014flexible,axenie2016self}.
Temporal correlation learning rules, such as Hebbian learning, shape the dynamics and structure of neural networks, as the study of \cite{siri2008mathematical} shows. The study provides excellent insights into effects involving a complex coupling between neuronal dynamics and synaptic graph structure that introduce both a structural and a dynamical point of view on neural network evolution. These principles are further expanded in the computational studies of \cite{cook2010unsupervised}, \cite{weber2007self}, \cite{firouzi2014flexible}, and \cite{axenie2016self}. Here, the interplay of HAR, WTA, and HL demonstrates how nonlinear sensory-motor correlations are extracted unsupervised from the noisy input streams. Additionally, efferent motor copies are used to generate plausible control inputs based solely on the sensory input and learnt sensorimotor correlations, as shown in the study of \cite{axenie2016self}.
\begin{figure}
    \centering
    \includegraphics[width=0.85\linewidth]{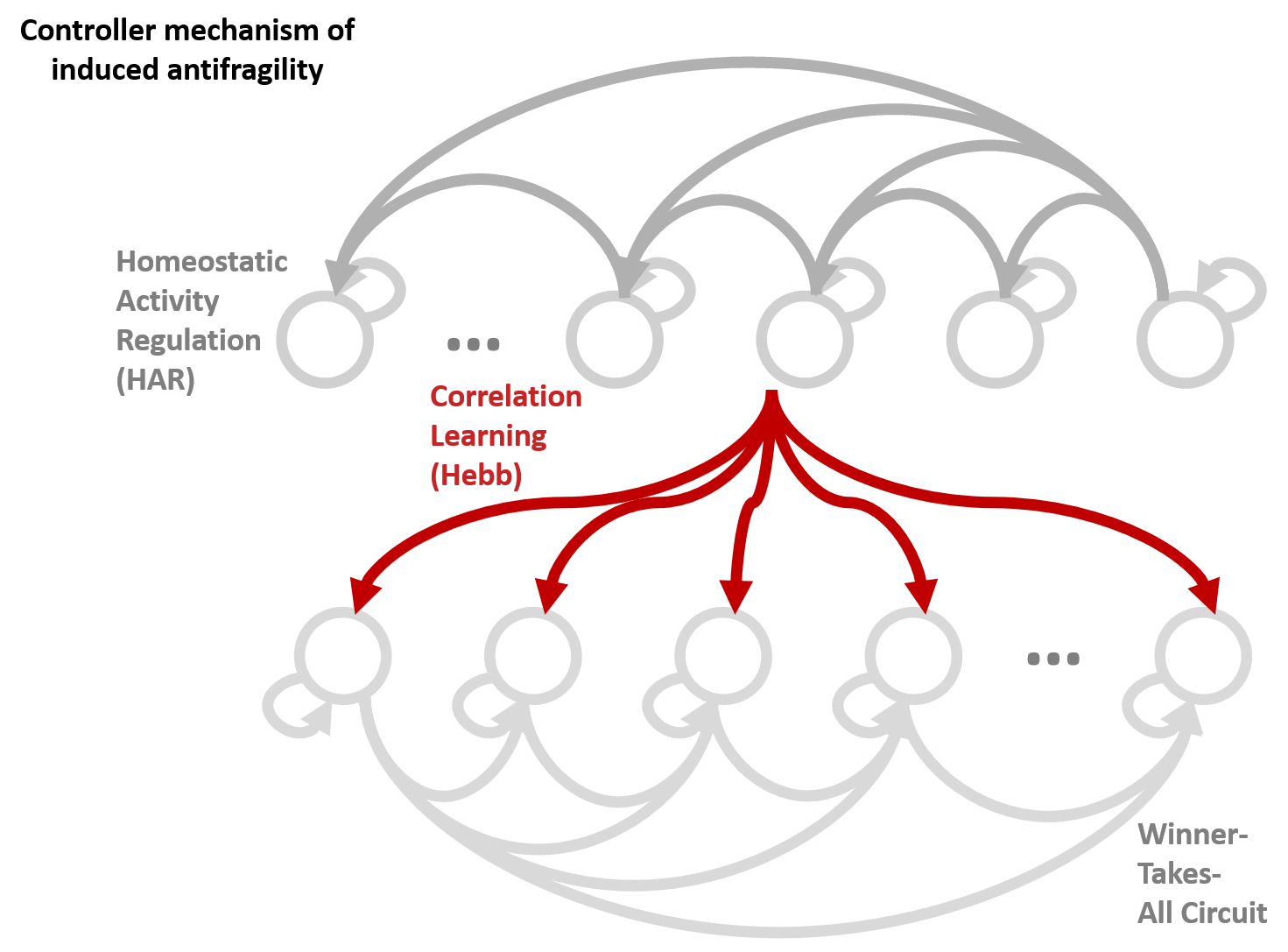}
    \caption{Induced antifragility core based on Hebbian Correlation Learning dynamics that capture and exploit temporal correlation of the sensorimotor input streams coded in the two interacting neuronal populations coding for sensory or motor quantities.}
    \label{fig:induced-antifragility-core}
\end{figure}
This prescribed dynamics-guided interplay of HAR, WTA, and HL is exploited in \cite{cook2010unsupervised} to demonstrate cue integration, inference, de-noising, and decision-making in sensorimotor representations within a single neural network. This is further extended to radial basis functions in \cite{firouzi2014flexible}, and reliability-modulated tuning curves in \cite{axenie2016self}. Here, attractor dynamics compute optimal output signals that enable tracking of the prescribed dynamics between two neuronal populations based on attractor dynamics and temporal correlation learning. These computational studies were also validated in closed-loop robotic implementations in \cite{weber2007self} and \cite{axenie2016self}, demonstrating the powerful transfer capacity of the antifragility concepts beyond analysis and modelling.
In this context, we postulate that antifragility is achieved through the interplay of HAR, WTA, and HL, as shown in Figure~\ref{fig:induced-antifragility-core}. There, time scale separation enables consistent adaptation to local changes, competition and cooperation ensure fast convergence and capacity building to anticipate changes in input data distribution, whereas correlation learning enables fast convergence towards the prescribed dynamics.
\begin{figure}[H]
    \centering
    \includegraphics[width=0.8\linewidth]{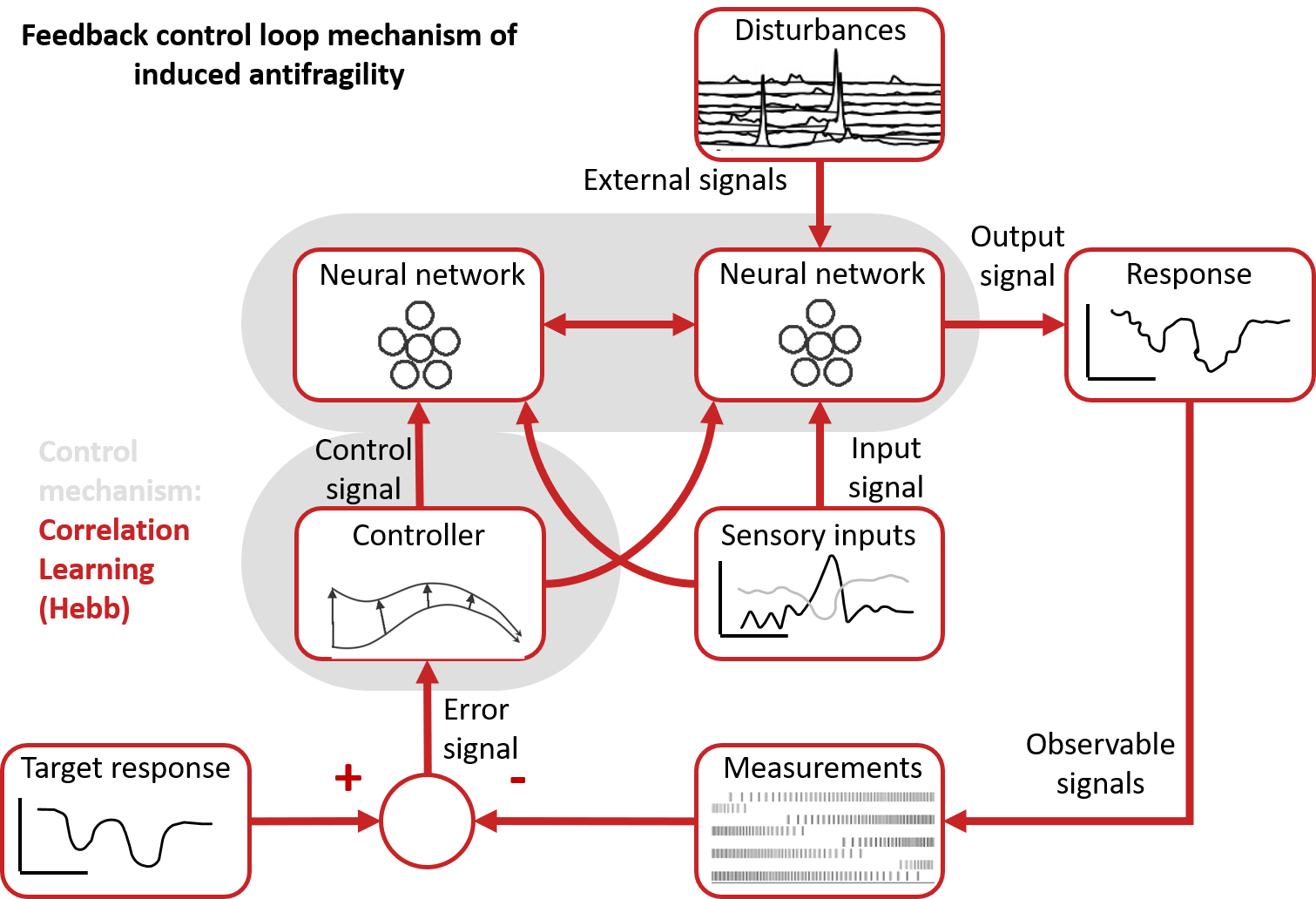}
    \caption{Induced antifragility: feedback control loop for induced antifragility components. The Controller implements computational mechanisms specific to between neural population dynamics of learning and integration (i.e. Hebbian Correlation Learning) and its interaction within and between populations' dynamics.}
    \label{fig:induced-antifragility}
\end{figure}

Our perspective complements the current tendencies in neuronal control systems where the event-based control paradigm unfolds over nested timescales \cite{sepulchre2022spiking}. Here, positive feedback is responsible for enhancing or amplifying change, whereas negative feedback dampens and buffers changes. The resulting mixed (positive/negative) controller acts as a monotone operator, which shapes the sensitivity of the closed-loop system and its excitability. Our paradigm goes away from typical equilibrium designs and describes behaviours which go beyond robustness.

\section{Outlook}

In our current study, we introduce the novel concept of neuronal processing antifragility in the context of sensorimotor control. In this preliminary work, we demonstrate that time scale harmonization can drive a sensorimotor network's evolution trajectory, through a judicious time scale separation within HAR, WTA, and HL. Having a network capable of structure variability through line attractors enables inference/learning relations among sensorimotor streams, de-noising, fusion, and decision-making. Finally, redundant overcompensation through the modulation of the closed-loop system builds capacity (i.e. redundant overcompensation) and ensures a fast reaching of the desired dynamics.
The purpose of this study is to instigate the community to consider antifragility as a novel analysis and design paradigm, shedding new light on the "canonical" neuronal computation mechanisms and their translation in practical neuromorphic implementations, for instance through the powerful Neuromorphic Intermediate Representation \cite{pedersen2023neuromorphic}. We believe that antifragile analysis and design can provide a new fruitful research direction in both computational models and their practical instantiations in technical systems.


\begin{thebibliography}{72}
\ifx \bisbn   \undefined \def \bisbn  #1{ISBN #1}\fi
\ifx \binits  \undefined \def \binits#1{#1}\fi
\ifx \bauthor  \undefined \def \bauthor#1{#1}\fi
\ifx \batitle  \undefined \def \batitle#1{#1}\fi
\ifx \bjtitle  \undefined \def \bjtitle#1{#1}\fi
\ifx \bvolume  \undefined \def \bvolume#1{\textbf{#1}}\fi
\ifx \byear  \undefined \def \byear#1{#1}\fi
\ifx \bissue  \undefined \def \bissue#1{#1}\fi
\ifx \bfpage  \undefined \def \bfpage#1{#1}\fi
\ifx \blpage  \undefined \def \blpage #1{#1}\fi
\ifx \burl  \undefined \def \burl#1{\textsf{#1}}\fi
\ifx \doiurl  \undefined \def \doiurl#1{\url{https://doi.org/#1}}\fi
\ifx \betal  \undefined \def \betal{\textit{et al.}}\fi
\ifx \binstitute  \undefined \def \binstitute#1{#1}\fi
\ifx \binstitutionaled  \undefined \def \binstitutionaled#1{#1}\fi
\ifx \bctitle  \undefined \def \bctitle#1{#1}\fi
\ifx \beditor  \undefined \def \beditor#1{#1}\fi
\ifx \bpublisher  \undefined \def \bpublisher#1{#1}\fi
\ifx \bbtitle  \undefined \def \bbtitle#1{#1}\fi
\ifx \bedition  \undefined \def \bedition#1{#1}\fi
\ifx \bseriesno  \undefined \def \bseriesno#1{#1}\fi
\ifx \blocation  \undefined \def \blocation#1{#1}\fi
\ifx \bsertitle  \undefined \def \bsertitle#1{#1}\fi
\ifx \bsnm \undefined \def \bsnm#1{#1}\fi
\ifx \bsuffix \undefined \def \bsuffix#1{#1}\fi
\ifx \bparticle \undefined \def \bparticle#1{#1}\fi
\ifx \barticle \undefined \def \barticle#1{#1}\fi
\bibcommenthead
\ifx \bconfdate \undefined \def \bconfdate #1{#1}\fi
\ifx \botherref \undefined \def \botherref #1{#1}\fi
\ifx \url \undefined \def \url#1{\textsf{#1}}\fi
\ifx \bchapter \undefined \def \bchapter#1{#1}\fi
\ifx \bbook \undefined \def \bbook#1{#1}\fi
\ifx \bcomment \undefined \def \bcomment#1{#1}\fi
\ifx \oauthor \undefined \def \oauthor#1{#1}\fi
\ifx \citeauthoryear \undefined \def \citeauthoryear#1{#1}\fi
\ifx \endbibitem  \undefined \def \endbibitem {}\fi
\ifx \bconflocation  \undefined \def \bconflocation#1{#1}\fi
\ifx \arxivurl  \undefined \def \arxivurl#1{\textsf{#1}}\fi
\csname PreBibitemsHook\endcsname

\bibitem[\protect\citeauthoryear{Payzan-LeNestour
  et~al.}{2013}]{payzan2013neural}
\begin{barticle}
\bauthor{\bsnm{Payzan-LeNestour}, \binits{E.}},
\bauthor{\bsnm{Dunne}, \binits{S.}},
\bauthor{\bsnm{Bossaerts}, \binits{P.}},
\bauthor{\bsnm{O’Doherty}, \binits{J.P.}}:
\batitle{The neural representation of unexpected uncertainty during value-based
  decision making}.
\bjtitle{Neuron}
\bvolume{79}(\bissue{1}),
\bfpage{191}--\blpage{201}
(\byear{2013})
\end{barticle}
\endbibitem

\bibitem[\protect\citeauthoryear{Findling et~al.}{2021}]{findling2021imprecise}
\begin{barticle}
\bauthor{\bsnm{Findling}, \binits{C.}},
\bauthor{\bsnm{Chopin}, \binits{N.}},
\bauthor{\bsnm{Koechlin}, \binits{E.}}:
\batitle{Imprecise neural computations as a source of adaptive behaviour in
  volatile environments}.
\bjtitle{Nature Human Behaviour}
\bvolume{5}(\bissue{1}),
\bfpage{99}--\blpage{112}
(\byear{2021})
\end{barticle}
\endbibitem

\bibitem[\protect\citeauthoryear{Cockburn et~al.}{2021}]{cockburn2021novelty}
\begin{botherref}
\oauthor{\bsnm{Cockburn}, \binits{J.}},
\oauthor{\bsnm{Man}, \binits{V.}},
\oauthor{\bsnm{Cunningham}, \binits{W.}},
\oauthor{\bsnm{O’Doherty}, \binits{J.P.}}:
Novelty and uncertainty interact to regulate the balance between exploration
  and exploitation in the human brain.
bioRxiv,
2021--10
(2021)
\end{botherref}
\endbibitem

\bibitem[\protect\citeauthoryear{Wu et~al.}{2021}]{wu2021better}
\begin{barticle}
\bauthor{\bsnm{Wu}, \binits{S.}},
\bauthor{\bsnm{Sun}, \binits{S.}},
\bauthor{\bsnm{Camilleri}, \binits{J.A.}},
\bauthor{\bsnm{Eickhoff}, \binits{S.B.}},
\bauthor{\bsnm{Yu}, \binits{R.}}:
\batitle{Better the devil you know than the devil you don't: Neural processing
  of risk and ambiguity}.
\bjtitle{NeuroImage}
\bvolume{236},
\bfpage{118109}
(\byear{2021})
\end{barticle}
\endbibitem

\bibitem[\protect\citeauthoryear{Iglesias
  et~al.}{2021}]{iglesias2021cholinergic}
\begin{barticle}
\bauthor{\bsnm{Iglesias}, \binits{S.}},
\bauthor{\bsnm{Kasper}, \binits{L.}},
\bauthor{\bsnm{Harrison}, \binits{S.J.}},
\bauthor{\bsnm{Manka}, \binits{R.}},
\bauthor{\bsnm{Mathys}, \binits{C.}},
\bauthor{\bsnm{Stephan}, \binits{K.E.}}:
\batitle{Cholinergic and dopaminergic effects on prediction error and
  uncertainty responses during sensory associative learning}.
\bjtitle{NeuroImage}
\bvolume{226},
\bfpage{117590}
(\byear{2021})
\end{barticle}
\endbibitem

\bibitem[\protect\citeauthoryear{Soltani and
  Izquierdo}{2019}]{soltani2019adaptive}
\begin{barticle}
\bauthor{\bsnm{Soltani}, \binits{A.}},
\bauthor{\bsnm{Izquierdo}, \binits{A.}}:
\batitle{Adaptive learning under expected and unexpected uncertainty}.
\bjtitle{Nature Reviews Neuroscience}
\bvolume{20}(\bissue{10}),
\bfpage{635}--\blpage{644}
(\byear{2019})
\end{barticle}
\endbibitem

\bibitem[\protect\citeauthoryear{Grossman et~al.}{2022}]{grossman2022serotonin}
\begin{barticle}
\bauthor{\bsnm{Grossman}, \binits{C.D.}},
\bauthor{\bsnm{Bari}, \binits{B.A.}},
\bauthor{\bsnm{Cohen}, \binits{J.Y.}}:
\batitle{Serotonin neurons modulate learning rate through uncertainty}.
\bjtitle{Current Biology}
\bvolume{32}(\bissue{3}),
\bfpage{586}--\blpage{599}
(\byear{2022})
\end{barticle}
\endbibitem

\bibitem[\protect\citeauthoryear{Bach et~al.}{2011}]{bach2011known}
\begin{barticle}
\bauthor{\bsnm{Bach}, \binits{D.R.}},
\bauthor{\bsnm{Hulme}, \binits{O.}},
\bauthor{\bsnm{Penny}, \binits{W.D.}},
\bauthor{\bsnm{Dolan}, \binits{R.J.}}:
\batitle{The known unknowns: neural representation of second-order uncertainty,
  and ambiguity}.
\bjtitle{Journal of Neuroscience}
\bvolume{31}(\bissue{13}),
\bfpage{4811}--\blpage{4820}
(\byear{2011})
\end{barticle}
\endbibitem

\bibitem[\protect\citeauthoryear{Van~Bergen and
  Jehee}{2019}]{van2019probabilistic}
\begin{barticle}
\bauthor{\bsnm{Van~Bergen}, \binits{R.S.}},
\bauthor{\bsnm{Jehee}, \binits{J.F.}}:
\batitle{Probabilistic representation in human visual cortex reflects
  uncertainty in serial decisions}.
\bjtitle{Journal of Neuroscience}
\bvolume{39}(\bissue{41}),
\bfpage{8164}--\blpage{8176}
(\byear{2019})
\end{barticle}
\endbibitem

\bibitem[\protect\citeauthoryear{Ma and Jazayeri}{2014}]{ma2014neural}
\begin{barticle}
\bauthor{\bsnm{Ma}, \binits{W.J.}},
\bauthor{\bsnm{Jazayeri}, \binits{M.}}:
\batitle{Neural coding of uncertainty and probability}.
\bjtitle{Annual review of neuroscience}
\bvolume{37},
\bfpage{205}--\blpage{220}
(\byear{2014})
\end{barticle}
\endbibitem

\bibitem[\protect\citeauthoryear{Bach and Dolan}{2012}]{bach2012knowing}
\begin{barticle}
\bauthor{\bsnm{Bach}, \binits{D.R.}},
\bauthor{\bsnm{Dolan}, \binits{R.J.}}:
\batitle{Knowing how much you don't know: a neural organization of uncertainty
  estimates}.
\bjtitle{Nature reviews neuroscience}
\bvolume{13}(\bissue{8}),
\bfpage{572}--\blpage{586}
(\byear{2012})
\end{barticle}
\endbibitem

\bibitem[\protect\citeauthoryear{Muller et~al.}{2019}]{muller2019control}
\begin{barticle}
\bauthor{\bsnm{Muller}, \binits{T.H.}},
\bauthor{\bsnm{Mars}, \binits{R.B.}},
\bauthor{\bsnm{Behrens}, \binits{T.E.}},
\bauthor{\bsnm{O'Reilly}, \binits{J.X.}}:
\batitle{Control of entropy in neural models of environmental state}.
\bjtitle{elife}
\bvolume{8},
\bfpage{39404}
(\byear{2019})
\end{barticle}
\endbibitem

\bibitem[\protect\citeauthoryear{Friston et~al.}{2012}]{friston2012dopamine}
\begin{barticle}
\bauthor{\bsnm{Friston}, \binits{K.J.}},
\bauthor{\bsnm{Shiner}, \binits{T.}},
\bauthor{\bsnm{FitzGerald}, \binits{T.}},
\bauthor{\bsnm{Galea}, \binits{J.M.}},
\bauthor{\bsnm{Adams}, \binits{R.}},
\bauthor{\bsnm{Brown}, \binits{H.}},
\bauthor{\bsnm{Dolan}, \binits{R.J.}},
\bauthor{\bsnm{Moran}, \binits{R.}},
\bauthor{\bsnm{Stephan}, \binits{K.E.}},
\bauthor{\bsnm{Bestmann}, \binits{S.}}:
\batitle{Dopamine, affordance and active inference}.
\bjtitle{PLoS computational biology}
\bvolume{8}(\bissue{1}),
\bfpage{1002327}
(\byear{2012})
\end{barticle}
\endbibitem

\bibitem[\protect\citeauthoryear{Feldman and
  Friston}{2010}]{feldman2010attention}
\begin{barticle}
\bauthor{\bsnm{Feldman}, \binits{H.}},
\bauthor{\bsnm{Friston}, \binits{K.J.}}:
\batitle{Attention, uncertainty, and free-energy}.
\bjtitle{Frontiers in human neuroscience}
\bvolume{4},
\bfpage{215}
(\byear{2010})
\end{barticle}
\endbibitem

\bibitem[\protect\citeauthoryear{Bland and Schaefer}{2012}]{bland2012different}
\begin{barticle}
\bauthor{\bsnm{Bland}, \binits{A.R.}},
\bauthor{\bsnm{Schaefer}, \binits{A.}}:
\batitle{Different varieties of uncertainty in human decision-making}.
\bjtitle{Frontiers in neuroscience}
\bvolume{6},
\bfpage{85}
(\byear{2012})
\end{barticle}
\endbibitem

\bibitem[\protect\citeauthoryear{Mushtaq et~al.}{2011}]{mushtaq2011uncertainty}
\begin{barticle}
\bauthor{\bsnm{Mushtaq}, \binits{F.}},
\bauthor{\bsnm{Bland}, \binits{A.R.}},
\bauthor{\bsnm{Schaefer}, \binits{A.}}:
\batitle{Uncertainty and cognitive control}.
\bjtitle{Frontiers in psychology}
\bvolume{2},
\bfpage{249}
(\byear{2011})
\end{barticle}
\endbibitem

\bibitem[\protect\citeauthoryear{Monosov}{2020}]{monosov2020outcome}
\begin{barticle}
\bauthor{\bsnm{Monosov}, \binits{I.E.}}:
\batitle{How outcome uncertainty mediates attention, learning, and
  decision-making}.
\bjtitle{Trends in neurosciences}
\bvolume{43}(\bissue{10}),
\bfpage{795}--\blpage{809}
(\byear{2020})
\end{barticle}
\endbibitem

\bibitem[\protect\citeauthoryear{Angela and
  Dayan}{2005}]{angela2005uncertainty}
\begin{barticle}
\bauthor{\bsnm{Angela}, \binits{J.Y.}},
\bauthor{\bsnm{Dayan}, \binits{P.}}:
\batitle{Uncertainty, neuromodulation, and attention}.
\bjtitle{Neuron}
\bvolume{46}(\bissue{4}),
\bfpage{681}--\blpage{692}
(\byear{2005})
\end{barticle}
\endbibitem

\bibitem[\protect\citeauthoryear{Schultz et~al.}{2008}]{schultz2008explicit}
\begin{barticle}
\bauthor{\bsnm{Schultz}, \binits{W.}},
\bauthor{\bsnm{Preuschoff}, \binits{K.}},
\bauthor{\bsnm{Camerer}, \binits{C.}},
\bauthor{\bsnm{Hsu}, \binits{M.}},
\bauthor{\bsnm{Fiorillo}, \binits{C.D.}},
\bauthor{\bsnm{Tobler}, \binits{P.N.}},
\bauthor{\bsnm{Bossaerts}, \binits{P.}}:
\batitle{Explicit neural signals reflecting reward uncertainty}.
\bjtitle{Philosophical Transactions of the Royal Society B: Biological
  Sciences}
\bvolume{363}(\bissue{1511}),
\bfpage{3801}--\blpage{3811}
(\byear{2008})
\end{barticle}
\endbibitem

\bibitem[\protect\citeauthoryear{Kosciessa
  et~al.}{2021}]{kosciessa2021thalamocortical}
\begin{barticle}
\bauthor{\bsnm{Kosciessa}, \binits{J.Q.}},
\bauthor{\bsnm{Lindenberger}, \binits{U.}},
\bauthor{\bsnm{Garrett}, \binits{D.D.}}:
\batitle{Thalamocortical excitability modulation guides human perception under
  uncertainty}.
\bjtitle{Nature Communications}
\bvolume{12}(\bissue{1}),
\bfpage{2430}
(\byear{2021})
\end{barticle}
\endbibitem

\bibitem[\protect\citeauthoryear{Cannon}{1929}]{cannon1929organization}
\begin{barticle}
\bauthor{\bsnm{Cannon}, \binits{W.B.}}:
\batitle{Organization for physiological homeostasis}.
\bjtitle{Physiological reviews}
\bvolume{9}(\bissue{3}),
\bfpage{399}--\blpage{431}
(\byear{1929})
\end{barticle}
\endbibitem

\bibitem[\protect\citeauthoryear{Musslick et~al.}{2019}]{musslick2019stability}
\begin{bchapter}
\bauthor{\bsnm{Musslick}, \binits{S.}},
\bauthor{\bsnm{Bizyaeva}, \binits{A.}},
\bauthor{\bsnm{Agaron}, \binits{S.}},
\bauthor{\bsnm{Leonard}, \binits{N.}},
\bauthor{\bsnm{Cohen}, \binits{J.D.}}:
\bctitle{Stability-flexibility dilemma in cognitive control: A dynamical system
  perspective}.
In: \bbtitle{Proceedings of the 41st Annual Meeting of the Cognitive Science
  Society}
(\byear{2019})
\end{bchapter}
\endbibitem

\bibitem[\protect\citeauthoryear{Holling}{1973}]{holling1973resilience}
\begin{barticle}
\bauthor{\bsnm{Holling}, \binits{C.S.}}:
\batitle{Resilience and stability of ecological systems}.
\bjtitle{Annual review of ecology and systematics}
\bvolume{4}(\bissue{1}),
\bfpage{1}--\blpage{23}
(\byear{1973})
\end{barticle}
\endbibitem

\bibitem[\protect\citeauthoryear{Braun}{2015}]{braun2015unforeseen}
\begin{barticle}
\bauthor{\bsnm{Braun}, \binits{E.}}:
\batitle{The unforeseen challenge: from genotype-to-phenotype in cell
  populations}.
\bjtitle{Reports on Progress in Physics}
\bvolume{78}(\bissue{3}),
\bfpage{036602}
(\byear{2015})
\end{barticle}
\endbibitem

\bibitem[\protect\citeauthoryear{Krakovsk{\'a}
  et~al.}{2021}]{krakovska2021resilience}
\begin{botherref}
\oauthor{\bsnm{Krakovsk{\'a}}, \binits{H.}},
\oauthor{\bsnm{Kuehn}, \binits{C.}},
\oauthor{\bsnm{Longo}, \binits{I.P.}}:
Resilience of dynamical systems.
European Journal of Applied Mathematics,
1--46
(2021)
\end{botherref}
\endbibitem

\bibitem[\protect\citeauthoryear{Marom and Marder}{2023}]{marom2023biophysical}
\begin{barticle}
\bauthor{\bsnm{Marom}, \binits{S.}},
\bauthor{\bsnm{Marder}, \binits{E.}}:
\batitle{A biophysical perspective on the resilience of neuronal excitability
  across timescales}.
\bjtitle{Nature Reviews Neuroscience}
\bvolume{24}(\bissue{10}),
\bfpage{640}--\blpage{652}
(\byear{2023})
\end{barticle}
\endbibitem

\bibitem[\protect\citeauthoryear{Taleb}{2012}]{taleb2012Antifragile}
\begin{bbook}
\bauthor{\bsnm{Taleb}, \binits{N.N.}}:
\bbtitle{Antifragile: Things that Gain from Disorder}.
\bpublisher{Random House Incorporated},
\blocation{New York}
(\byear{2012})
\end{bbook}
\endbibitem

\bibitem[\protect\citeauthoryear{Taleb and
  Douady}{2013}]{taleb2013mathematical}
\begin{barticle}
\bauthor{\bsnm{Taleb}, \binits{N.N.}},
\bauthor{\bsnm{Douady}, \binits{R.}}:
\batitle{Mathematical definition, mapping, and detection of (anti) fragility}.
\bjtitle{Quantitative Finance}
\bvolume{13}(\bissue{11}),
\bfpage{1677}--\blpage{1689}
(\byear{2013})
\end{barticle}
\endbibitem

\bibitem[\protect\citeauthoryear{Meyers and Bull}{2002}]{meyers2002fighting}
\begin{barticle}
\bauthor{\bsnm{Meyers}, \binits{L.A.}},
\bauthor{\bsnm{Bull}, \binits{J.J.}}:
\batitle{Fighting change with change: adaptive variation in an uncertain
  world}.
\bjtitle{Trends in Ecology \& Evolution}
\bvolume{17}(\bissue{12}),
\bfpage{551}--\blpage{557}
(\byear{2002})
\end{barticle}
\endbibitem

\bibitem[\protect\citeauthoryear{Meyer}{2015}]{meyer2015dynamical}
\begin{botherref}
\oauthor{\bsnm{Meyer}, \binits{K.}}:
A dynamical systems framework for resilience in ecology.
arXiv preprint arXiv:1509.08175
(2015)
\end{botherref}
\endbibitem

\bibitem[\protect\citeauthoryear{Lara}{2018}]{lara2018mathematical}
\begin{barticle}
\bauthor{\bsnm{Lara}, \binits{M.D.}}:
\batitle{A mathematical framework for resilience: dynamics, uncertainties,
  strategies, and recovery regimes}.
\bjtitle{Environmental Modeling \& Assessment}
\bvolume{23}(\bissue{6}),
\bfpage{703}--\blpage{712}
(\byear{2018})
\end{barticle}
\endbibitem

\bibitem[\protect\citeauthoryear{Hebbar et~al.}{2022}]{hebbar2022interplay}
\begin{botherref}
\oauthor{\bsnm{Hebbar}, \binits{A.}},
\oauthor{\bsnm{Moger}, \binits{A.}},
\oauthor{\bsnm{Hari}, \binits{K.}},
\oauthor{\bsnm{Jolly}, \binits{M.K.}}:
Interplay of positive and negative feedback loops governs robustness in
  multistable biological networks.
bioRxiv
(2022)
\end{botherref}
\endbibitem

\bibitem[\protect\citeauthoryear{Bramson}{2010}]{bramson2010formal}
\begin{bchapter}
\bauthor{\bsnm{Bramson}, \binits{A.L.}}:
\bctitle{Formal measures of dynamical properties: robustness and
  sustainability}.
In: \bbtitle{2010 AAAI Fall Symposium Series}
(\byear{2010})
\end{bchapter}
\endbibitem

\bibitem[\protect\citeauthoryear{Balchanos}{2012}]{balchanos2012probabilistic}
\begin{bbook}
\bauthor{\bsnm{Balchanos}, \binits{M.G.}}:
\bbtitle{A Probabilistic Technique for the Assessment of Complex Dynamic System
  Resilience},
(\byear{2012})
\end{bbook}
\endbibitem

\bibitem[\protect\citeauthoryear{Arnoldi et~al.}{2016}]{arnoldi2016resilience}
\begin{barticle}
\bauthor{\bsnm{Arnoldi}, \binits{J.-F.}},
\bauthor{\bsnm{Loreau}, \binits{M.}},
\bauthor{\bsnm{Haegeman}, \binits{B.}}:
\batitle{Resilience, reactivity and variability: A mathematical comparison of
  ecological stability measures}.
\bjtitle{Journal of theoretical biology}
\bvolume{389},
\bfpage{47}--\blpage{59}
(\byear{2016})
\end{barticle}
\endbibitem

\bibitem[\protect\citeauthoryear{Ay and Krakauer}{2007}]{ay2007geometric}
\begin{barticle}
\bauthor{\bsnm{Ay}, \binits{N.}},
\bauthor{\bsnm{Krakauer}, \binits{D.C.}}:
\batitle{Geometric robustness theory and biological networks}.
\bjtitle{Theory in biosciences}
\bvolume{125},
\bfpage{93}--\blpage{121}
(\byear{2007})
\end{barticle}
\endbibitem

\bibitem[\protect\citeauthoryear{Orb{\'a}n and
  Wolpert}{2011}]{orban2011representations}
\begin{barticle}
\bauthor{\bsnm{Orb{\'a}n}, \binits{G.}},
\bauthor{\bsnm{Wolpert}, \binits{D.M.}}:
\batitle{Representations of uncertainty in sensorimotor control}.
\bjtitle{Current opinion in neurobiology}
\bvolume{21}(\bissue{4}),
\bfpage{629}--\blpage{635}
(\byear{2011})
\end{barticle}
\endbibitem

\bibitem[\protect\citeauthoryear{Koblinger
  et~al.}{2021}]{koblinger2021representations}
\begin{barticle}
\bauthor{\bsnm{Koblinger}, \binits{{\'A}.}},
\bauthor{\bsnm{Fiser}, \binits{J.}},
\bauthor{\bsnm{Lengyel}, \binits{M.}}:
\batitle{Representations of uncertainty: where art thou?}
\bjtitle{Current Opinion in Behavioral Sciences}
\bvolume{38},
\bfpage{150}--\blpage{162}
(\byear{2021})
\end{barticle}
\endbibitem

\bibitem[\protect\citeauthoryear{K{\"o}rding and
  Wolpert}{2004}]{kording2004bayesian}
\begin{barticle}
\bauthor{\bsnm{K{\"o}rding}, \binits{K.P.}},
\bauthor{\bsnm{Wolpert}, \binits{D.M.}}:
\batitle{Bayesian integration in sensorimotor learning}.
\bjtitle{Nature}
\bvolume{427}(\bissue{6971}),
\bfpage{244}--\blpage{247}
(\byear{2004})
\end{barticle}
\endbibitem

\bibitem[\protect\citeauthoryear{Trommersh{\"a}user}{2009}]{trommershauser2009biases}
\begin{barticle}
\bauthor{\bsnm{Trommersh{\"a}user}, \binits{J.}}:
\batitle{Biases and optimality of sensory-motor and cognitive decisions}.
\bjtitle{Progress in brain research}
\bvolume{174},
\bfpage{267}--\blpage{278}
(\byear{2009})
\end{barticle}
\endbibitem

\bibitem[\protect\citeauthoryear{Trommersh{\"a}user
  et~al.}{2008}]{trommershauser2008decision}
\begin{barticle}
\bauthor{\bsnm{Trommersh{\"a}user}, \binits{J.}},
\bauthor{\bsnm{Maloney}, \binits{L.T.}},
\bauthor{\bsnm{Landy}, \binits{M.S.}}:
\batitle{Decision making, movement planning and statistical decision theory}.
\bjtitle{Trends in cognitive sciences}
\bvolume{12}(\bissue{8}),
\bfpage{291}--\blpage{297}
(\byear{2008})
\end{barticle}
\endbibitem

\bibitem[\protect\citeauthoryear{Taleb}{2020}]{taleb2020statistical}
\begin{botherref}
\oauthor{\bsnm{Taleb}, \binits{N.N.}}:
Statistical consequences of fat tails: Real world preasymptotics, epistemology,
  and applications.
arXiv preprint arXiv:2001.10488
(2020)
\end{botherref}
\endbibitem

\bibitem[\protect\citeauthoryear{van Beers et~al.}{2002}]{van2002role}
\begin{barticle}
\bauthor{\bsnm{Beers}, \binits{R.J.}},
\bauthor{\bsnm{Baraduc}, \binits{P.}},
\bauthor{\bsnm{Wolpert}, \binits{D.M.}}:
\batitle{Role of uncertainty in sensorimotor control}.
\bjtitle{Philosophical Transactions of the Royal Society of London. Series B:
  Biological Sciences}
\bvolume{357}(\bissue{1424}),
\bfpage{1137}--\blpage{1145}
(\byear{2002})
\end{barticle}
\endbibitem

\bibitem[\protect\citeauthoryear{Chavez-Garcia
  et~al.}{2016}]{chavez2016discovering}
\begin{bchapter}
\bauthor{\bsnm{Chavez-Garcia}, \binits{R.O.}},
\bauthor{\bsnm{Luce-Vayrac}, \binits{P.}},
\bauthor{\bsnm{Chatila}, \binits{R.}}:
\bctitle{Discovering affordances through perception and manipulation}.
In: \bbtitle{2016 IEEE/RSJ International Conference on Intelligent Robots and
  Systems (IROS)},
pp. \bfpage{3959}--\blpage{3964}
(\byear{2016}).
\bcomment{IEEE}
\end{bchapter}
\endbibitem

\bibitem[\protect\citeauthoryear{Ogawa et~al.}{2006}]{ogawa2006adaptive}
\begin{barticle}
\bauthor{\bsnm{Ogawa}, \binits{N.}},
\bauthor{\bsnm{Sakaguchi}, \binits{Y.}},
\bauthor{\bsnm{Namiki}, \binits{A.}},
\bauthor{\bsnm{Ishikawa}, \binits{M.}}:
\batitle{Adaptive acquisition of dynamics matching in sensory-motor fusion
  system}.
\bjtitle{Electronics and Communications in Japan (Part III: Fundamental
  Electronic Science)}
\bvolume{89}(\bissue{7}),
\bfpage{19}--\blpage{30}
(\byear{2006})
\end{barticle}
\endbibitem

\bibitem[\protect\citeauthoryear{Fiser et~al.}{2010}]{fiser2010statistically}
\begin{barticle}
\bauthor{\bsnm{Fiser}, \binits{J.}},
\bauthor{\bsnm{Berkes}, \binits{P.}},
\bauthor{\bsnm{Orb{\'a}n}, \binits{G.}},
\bauthor{\bsnm{Lengyel}, \binits{M.}}:
\batitle{Statistically optimal perception and learning: from behavior to neural
  representations}.
\bjtitle{Trends in cognitive sciences}
\bvolume{14}(\bissue{3}),
\bfpage{119}--\blpage{130}
(\byear{2010})
\end{barticle}
\endbibitem

\bibitem[\protect\citeauthoryear{Ghahramani
  et~al.}{1997}]{ghahramani1997computational}
\begin{botherref}
\oauthor{\bsnm{Ghahramani}, \binits{Z.}},
\oauthor{\bsnm{Wolptrt}, \binits{D.M.}},
\oauthor{\bsnm{Jordan}, \binits{M.I.}}:
Computational models of sensorimotor integration
\textbf{119},
117--147
(1997)
\end{botherref}
\endbibitem

\bibitem[\protect\citeauthoryear{Nagata et~al.}{1994}]{nagata1994hierarchical}
\begin{barticle}
\bauthor{\bsnm{Nagata}, \binits{S.}},
\bauthor{\bsnm{Masumoto}, \binits{D.}},
\bauthor{\bsnm{Yamakawa}, \binits{H.}},
\bauthor{\bsnm{Kimoto}, \binits{T.}}:
\batitle{Hierarchical sensory-motor fusion model with neural networks}.
\bjtitle{Journal of the Robotics Society of Japan}
\bvolume{12}(\bissue{5}),
\bfpage{685}--\blpage{694}
(\byear{1994})
\end{barticle}
\endbibitem

\bibitem[\protect\citeauthoryear{Schlicht and
  Schrater}{2007}]{schlicht2007impact}
\begin{barticle}
\bauthor{\bsnm{Schlicht}, \binits{E.J.}},
\bauthor{\bsnm{Schrater}, \binits{P.R.}}:
\batitle{Impact of coordinate transformation uncertainty on human sensorimotor
  control}.
\bjtitle{Journal of neurophysiology}
\bvolume{97}(\bissue{6}),
\bfpage{4203}--\blpage{4214}
(\byear{2007})
\end{barticle}
\endbibitem

\bibitem[\protect\citeauthoryear{Knill and Pouget}{2004}]{knill2004bayesian}
\begin{barticle}
\bauthor{\bsnm{Knill}, \binits{D.C.}},
\bauthor{\bsnm{Pouget}, \binits{A.}}:
\batitle{The bayesian brain: the role of uncertainty in neural coding and
  computation}.
\bjtitle{TRENDS in Neurosciences}
\bvolume{27}(\bissue{12}),
\bfpage{712}--\blpage{719}
(\byear{2004})
\end{barticle}
\endbibitem

\bibitem[\protect\citeauthoryear{Berniker and
  Kording}{2011}]{berniker2011bayesian}
\begin{barticle}
\bauthor{\bsnm{Berniker}, \binits{M.}},
\bauthor{\bsnm{Kording}, \binits{K.}}:
\batitle{Bayesian approaches to sensory integration for motor control}.
\bjtitle{Wiley Interdisciplinary Reviews: Cognitive Science}
\bvolume{2}(\bissue{4}),
\bfpage{419}--\blpage{428}
(\byear{2011})
\end{barticle}
\endbibitem

\bibitem[\protect\citeauthoryear{Topel et~al.}{2023}]{topel2023expecting}
\begin{botherref}
\oauthor{\bsnm{Topel}, \binits{S.}},
\oauthor{\bsnm{Ma}, \binits{I.}},
\oauthor{\bsnm{Sleutels}, \binits{J.}},
\oauthor{\bsnm{Steenbergen}, \binits{H.}},
\oauthor{\bsnm{Bruijn}, \binits{E.R.}},
\oauthor{\bsnm{Duijvenvoorde}, \binits{A.C.}}:
Expecting the unexpected: a review of learning under uncertainty across
  development.
Cognitive, Affective, \& Behavioral Neuroscience,
1--21
(2023)
\end{botherref}
\endbibitem

\bibitem[\protect\citeauthoryear{Axenie et~al.}{2023}]{axenie2023antifragility}
\begin{botherref}
\oauthor{\bsnm{Axenie}, \binits{C.}},
\oauthor{\bsnm{López-Corona}, \binits{O.}},
\oauthor{\bsnm{Makridis}, \binits{M.A.}},
\oauthor{\bsnm{Akbarzadeh}, \binits{M.}},
\oauthor{\bsnm{Saveriano}, \binits{M.}},
\oauthor{\bsnm{Stancu}, \binits{A.}},
\oauthor{\bsnm{West}, \binits{J.}}:
Antifragility as a complex system's response to perturbations, volatility, and
  time
(2023)
\end{botherref}
\endbibitem

\bibitem[\protect\citeauthoryear{Axenie and Saveriano}{2023}]{10345540}
\begin{barticle}
\bauthor{\bsnm{Axenie}, \binits{C.}},
\bauthor{\bsnm{Saveriano}, \binits{M.}}:
\batitle{Antifragile control systems: The case of mobile robot trajectory
  tracking under uncertainty and volatility}.
\bjtitle{IEEE Access}
\bvolume{11},
\bfpage{138188}--\blpage{138200}
(\byear{2023})
\end{barticle}
\endbibitem

\bibitem[\protect\citeauthoryear{Axenie and
  Grossi}{2023}]{axenie2023antifragile}
\begin{botherref}
\oauthor{\bsnm{Axenie}, \binits{C.}},
\oauthor{\bsnm{Grossi}, \binits{M.}}:
Antifragile Control Systems: The case of an oscillator-based network model of
  urban road traffic dynamics
(2023)
\end{botherref}
\endbibitem

\bibitem[\protect\citeauthoryear{Axenie et~al.}{2022}]{axenie2022antifragile}
\begin{barticle}
\bauthor{\bsnm{Axenie}, \binits{C.}},
\bauthor{\bsnm{Kurz}, \binits{D.}},
\bauthor{\bsnm{Saveriano}, \binits{M.}}:
\batitle{Antifragile control systems: The case of an anti-symmetric network
  model of the tumor-immune-drug interactions}.
\bjtitle{Symmetry}
\bvolume{14}(\bissue{10}),
\bfpage{2034}
(\byear{2022})
\end{barticle}
\endbibitem

\bibitem[\protect\citeauthoryear{Pineda et~al.}{2019}]{pineda2019novel}
\begin{botherref}
\oauthor{\bsnm{Pineda}, \binits{O.K.}},
\oauthor{\bsnm{Kim}, \binits{H.}},
\oauthor{\bsnm{Gershenson}, \binits{C.}}, et al.:
A novel antifragility measure based on satisfaction and its application to
  random and biological boolean networks.
Complexity
\textbf{2019}
(2019)
\end{botherref}
\endbibitem

\bibitem[\protect\citeauthoryear{Kwon and Cho}{2008}]{kwon2008quantitative}
\begin{barticle}
\bauthor{\bsnm{Kwon}, \binits{Y.-K.}},
\bauthor{\bsnm{Cho}, \binits{K.-H.}}:
\batitle{Quantitative analysis of robustness and fragility in biological
  networks based on feedback dynamics}.
\bjtitle{Bioinformatics}
\bvolume{24}(\bissue{7}),
\bfpage{987}--\blpage{994}
(\byear{2008})
\end{barticle}
\endbibitem

\bibitem[\protect\citeauthoryear{Johnson and
  Gheorghe}{2013}]{johnson2013antifragility}
\begin{barticle}
\bauthor{\bsnm{Johnson}, \binits{J.}},
\bauthor{\bsnm{Gheorghe}, \binits{A.V.}}:
\batitle{Antifragility analysis and measurement framework for systems of
  systems}.
\bjtitle{International journal of disaster risk science}
\bvolume{4},
\bfpage{159}--\blpage{168}
(\byear{2013})
\end{barticle}
\endbibitem

\bibitem[\protect\citeauthoryear{de~Bruijn et~al.}{2020}]{de2020antifragility}
\begin{barticle}
\bauthor{\bsnm{Bruijn}, \binits{H.}},
\bauthor{\bsnm{Groessler}, \binits{A.}},
\bauthor{\bsnm{Videira}, \binits{N.}}:
\batitle{Antifragility as a design criterion for modelling dynamic systems}.
\bjtitle{Systems Research and Behavioral Science}
\bvolume{37}(\bissue{1}),
\bfpage{23}--\blpage{37}
(\byear{2020})
\end{barticle}
\endbibitem

\bibitem[\protect\citeauthoryear{Levin and Narendra}{1993}]{levin1993control}
\begin{barticle}
\bauthor{\bsnm{Levin}, \binits{A.U.}},
\bauthor{\bsnm{Narendra}, \binits{K.S.}}:
\batitle{Control of nonlinear dynamical systems using neural networks:
  Controllability and stabilization}.
\bjtitle{IEEE Transactions on neural networks}
\bvolume{4}(\bissue{2}),
\bfpage{192}--\blpage{206}
(\byear{1993})
\end{barticle}
\endbibitem

\bibitem[\protect\citeauthoryear{Levin and Narendra}{1996}]{levin1996control}
\begin{barticle}
\bauthor{\bsnm{Levin}, \binits{A.U.}},
\bauthor{\bsnm{Narendra}, \binits{K.S.}}:
\batitle{Control of nonlinear dynamical systems using neural networks. ii.
  observability, identification, and control}.
\bjtitle{IEEE transactions on neural networks}
\bvolume{7}(\bissue{1}),
\bfpage{30}--\blpage{42}
(\byear{1996})
\end{barticle}
\endbibitem

\bibitem[\protect\citeauthoryear{Narendra}{1996}]{narendra1996neural}
\begin{barticle}
\bauthor{\bsnm{Narendra}, \binits{K.S.}}:
\batitle{Neural networks for control theory and practice}.
\bjtitle{Proceedings of the IEEE}
\bvolume{84}(\bissue{10}),
\bfpage{1385}--\blpage{1406}
(\byear{1996})
\end{barticle}
\endbibitem

\bibitem[\protect\citeauthoryear{Cook et~al.}{2010}]{cook2010unsupervised}
\begin{bchapter}
\bauthor{\bsnm{Cook}, \binits{M.}},
\bauthor{\bsnm{Jug}, \binits{F.}},
\bauthor{\bsnm{Krautz}, \binits{C.}},
\bauthor{\bsnm{Steger}, \binits{A.}}:
\bctitle{Unsupervised learning of relations}.
In: \bbtitle{Artificial Neural Networks--ICANN 2010: 20th International
  Conference, Thessaloniki, Greece, September 15-18, 2010, Proceedings, Part I
  20},
pp. \bfpage{164}--\blpage{173}
(\byear{2010}).
\bcomment{Springer}
\end{bchapter}
\endbibitem

\bibitem[\protect\citeauthoryear{Weber and Wermter}{2007}]{weber2007self}
\begin{barticle}
\bauthor{\bsnm{Weber}, \binits{C.}},
\bauthor{\bsnm{Wermter}, \binits{S.}}:
\batitle{A self-organizing map of sigma--pi units}.
\bjtitle{Neurocomputing}
\bvolume{70}(\bissue{13-15}),
\bfpage{2552}--\blpage{2560}
(\byear{2007})
\end{barticle}
\endbibitem

\bibitem[\protect\citeauthoryear{Firouzi et~al.}{2014}]{firouzi2014flexible}
\begin{bchapter}
\bauthor{\bsnm{Firouzi}, \binits{M.}},
\bauthor{\bsnm{Glasauer}, \binits{S.}},
\bauthor{\bsnm{Conradt}, \binits{J.}}:
\bctitle{Flexible cue integration by line attraction dynamics and divisive
  normalization}.
In: \bbtitle{Artificial Neural Networks and Machine Learning--ICANN 2014: 24th
  International Conference on Artificial Neural Networks, Hamburg, Germany,
  September 15-19, 2014. Proceedings 24},
pp. \bfpage{691}--\blpage{698}
(\byear{2014}).
\bcomment{Springer}
\end{bchapter}
\endbibitem

\bibitem[\protect\citeauthoryear{Axenie et~al.}{2016}]{axenie2016self}
\begin{barticle}
\bauthor{\bsnm{Axenie}, \binits{C.}},
\bauthor{\bsnm{Richter}, \binits{C.}},
\bauthor{\bsnm{Conradt}, \binits{J.}}:
\batitle{A self-synthesis approach to perceptual learning for multisensory
  fusion in robotics}.
\bjtitle{Sensors}
\bvolume{16}(\bissue{10}),
\bfpage{1751}
(\byear{2016})
\end{barticle}
\endbibitem

\bibitem[\protect\citeauthoryear{Ruggiero
  et~al.}{2021}]{ruggiero2021mitochondria}
\begin{barticle}
\bauthor{\bsnm{Ruggiero}, \binits{A.}},
\bauthor{\bsnm{Katsenelson}, \binits{M.}},
\bauthor{\bsnm{Slutsky}, \binits{I.}}:
\batitle{Mitochondria: new players in homeostatic regulation of firing rate set
  points}.
\bjtitle{Trends in Neurosciences}
\bvolume{44}(\bissue{8}),
\bfpage{605}--\blpage{618}
(\byear{2021})
\end{barticle}
\endbibitem

\bibitem[\protect\citeauthoryear{Kim and Lim}{2022}]{kim2022dynamical}
\begin{barticle}
\bauthor{\bsnm{Kim}, \binits{S.-Y.}},
\bauthor{\bsnm{Lim}, \binits{W.}}:
\batitle{Dynamical origin for winner-take-all competition in a biological
  network of the hippocampal dentate gyrus}.
\bjtitle{Physical Review E}
\bvolume{105}(\bissue{1}),
\bfpage{014418}
(\byear{2022})
\end{barticle}
\endbibitem

\bibitem[\protect\citeauthoryear{Siri et~al.}{2008}]{siri2008mathematical}
\begin{barticle}
\bauthor{\bsnm{Siri}, \binits{B.}},
\bauthor{\bsnm{Berry}, \binits{H.}},
\bauthor{\bsnm{Cessac}, \binits{B.}},
\bauthor{\bsnm{Delord}, \binits{B.}},
\bauthor{\bsnm{Quoy}, \binits{M.}}:
\batitle{A mathematical analysis of the effects of hebbian learning rules on
  the dynamics and structure of discrete-time random recurrent neural
  networks}.
\bjtitle{Neural Computation}
\bvolume{20}(\bissue{12}),
\bfpage{2937}--\blpage{2966}
(\byear{2008})
\end{barticle}
\endbibitem

\bibitem[\protect\citeauthoryear{Sepulchre}{2022}]{sepulchre2022spiking}
\begin{barticle}
\bauthor{\bsnm{Sepulchre}, \binits{R.}}:
\batitle{Spiking control systems}.
\bjtitle{Proceedings of the IEEE}
\bvolume{110}(\bissue{5}),
\bfpage{577}--\blpage{589}
(\byear{2022})
\end{barticle}
\endbibitem

\bibitem[\protect\citeauthoryear{Pedersen
  et~al.}{2023}]{pedersen2023neuromorphic}
\begin{botherref}
\oauthor{\bsnm{Pedersen}, \binits{J.E.}},
\oauthor{\bsnm{Abreu}, \binits{S.}},
\oauthor{\bsnm{Jobst}, \binits{M.}},
\oauthor{\bsnm{Lenz}, \binits{G.}},
\oauthor{\bsnm{Fra}, \binits{V.}},
\oauthor{\bsnm{Bauer}, \binits{F.C.}},
\oauthor{\bsnm{Muir}, \binits{D.R.}},
\oauthor{\bsnm{Zhou}, \binits{P.}},
\oauthor{\bsnm{Vogginger}, \binits{B.}},
\oauthor{\bsnm{Heckel}, \binits{K.}}, et al.:
Neuromorphic intermediate representation: A unified instruction set for
  interoperable brain-inspired computing.
arXiv preprint arXiv:2311.14641
(2023)
\end{botherref}
\endbibitem

\end{thebibliography}


\end{document}